\documentclass[
  ,draft            
  ]
  {aipproc}

\layoutstyle{6x9}

\begin{document}

\title{Measuring dark matter by modeling interacting galaxies}


\author{Petsch, H.P.}{
  address={T\"urkenschanzstr. 17, 1180 Wien, Austria}
}

\author{R\accent23u\v{z}i\v{c}ka, A.}{
  address={T\"urkenschanzstr. 17, 1180 Wien, Austria}
}

\author{Theis, Ch.}{
  address={T\"urkenschanzstr. 17, 1180 Wien, Austria}
  ,altaddress={Planetarium Mannheim, 68165 Mannheim, Germany} 
}

\begin{abstract}
The dark matter content of galaxies is usually determined from galaxies in dynamical equilibrium, mainly from rotationally supported galactic components. Such determinations restrict measurements to special regions in galaxies, e.g.\,the galactic plane(s), whereas other regions are not probed at all. Interacting galaxies offer an alternative,
because extended tidal tails often probe outer or off-plane regions of galaxies. However, these systems are neither in dynamical equilibrium nor simple, because they are composed of two or more galaxies, by this increasing the associated parameter space.
We present our genetic algorithm based modeling tool which allows to investigate the extended parameter space of interacting galaxies. From these studies, we derive the dynamical history of (well observed) galaxies. Among other parameters we constrain the dark matter content of the involved galaxies. We demonstrate the applicability of this strategy with examples ranging from stellar streams around the Milky Way to extended tidal tails, from proto-typical binary galaxies (like M51 or the Antennae system) to small group of galaxies.

\end{abstract}

\maketitle


\section{Introduction}
Interacting galaxies often show extended arms and bridges, caused by the gravitational tides \cite{holmberg41}, \cite{pfleiderer61}, \cite{toomre72}. Modeling these interactions allows us to recover the kinematic history of observed galaxies and serves as a test bed for the physics of the merging process. Timescales, dynamical friction, galactic structures and many more can be investigated and verified by models.
By observing rotation curves of spiral galaxies, Zwicky \cite{zwicky37} has shown, that galaxies must contain an additional non visible mass. This dark matter (DM) was introduced to galaxy simulations mostly as a collision-less particle component. Shape and extension of the (DM-)halos have large effects on the behavior of interacting galaxies \cite{barnes92}.
State-of-the-art simulations include multiple components (stars, DM, different gas components, magnetic fields, radiation) and complex recipes for star formation and feedback, AGN influence etc. These simulations require a huge amount of computational power. For remodeling observed systems it is therefore important to know as many physical parameters of the interaction as possible in advance. To solve this problem, we coupled a fast restricted N-body code for galactic interactions to a genetic algorithm (GA). 

\begin{figure}
\centering
\rotatebox{-90}
{\includegraphics[bb=5 30 585 715, angle=90, scale=0.35, clip]{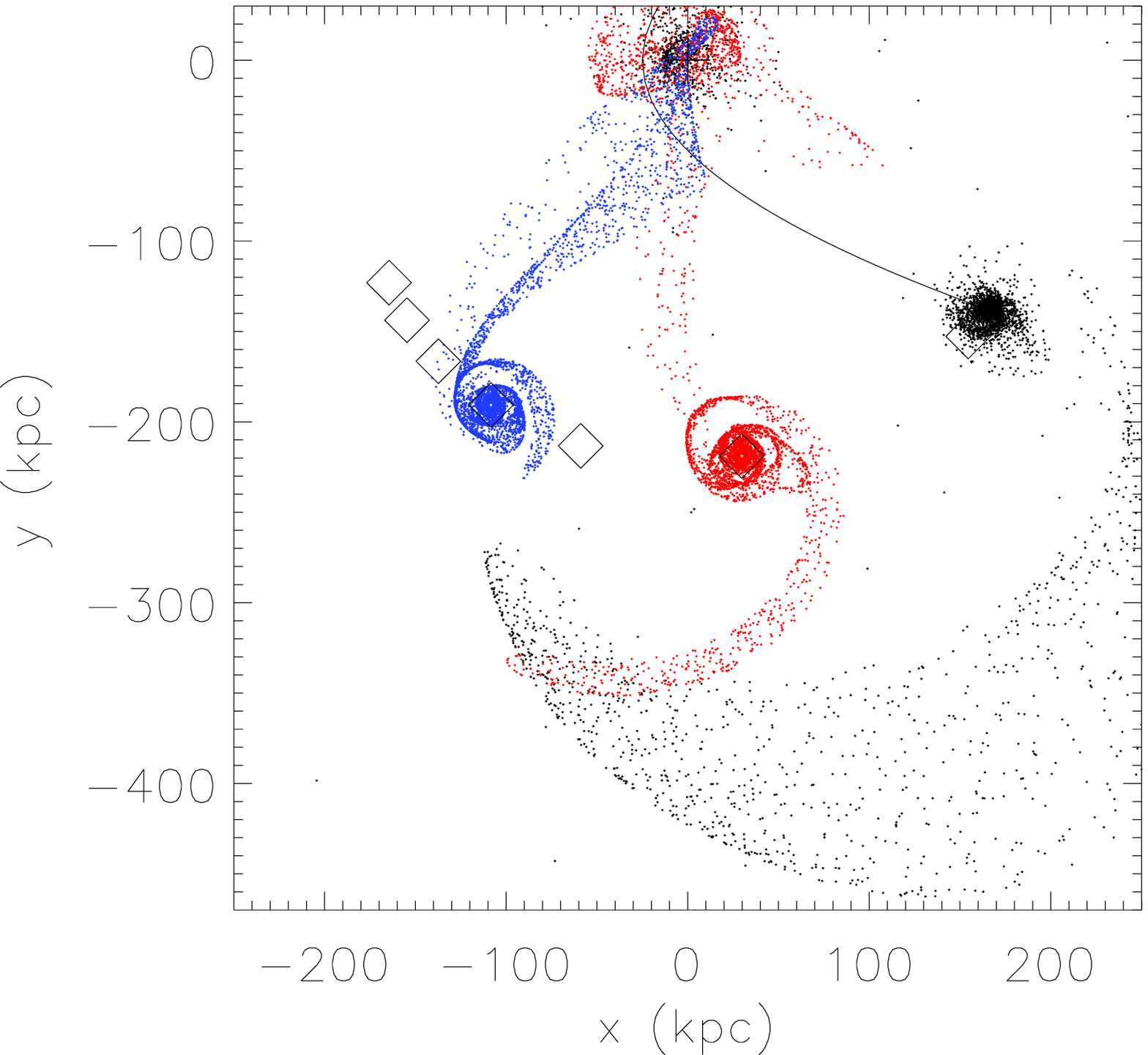}}
\rotatebox{-90}
{\includegraphics[bb=5 30 585 715, angle=90, scale=0.35, clip]{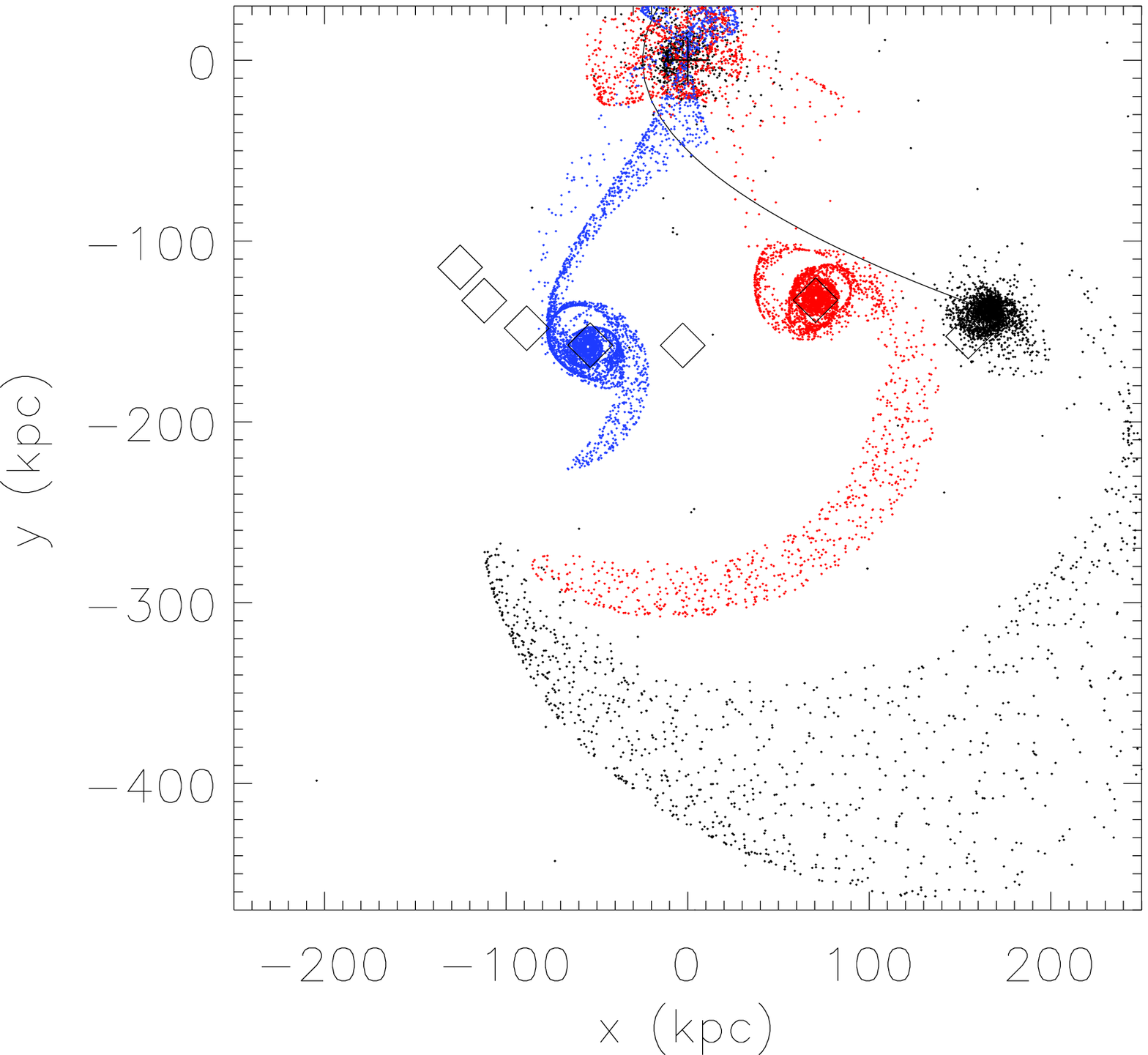}}

\caption{Left plot: Interaction of equal disk galaxies. Deflection from the parabolic orbit (for point masses) depending on (isothermal) halo size. From right to left \(r_\mathrm{halo}=0\)\,kpc to \(r_\mathrm{halo}=140\)\,kpc in steps of 20\,kpc (centers marked with diamonds). Right plot: same as left, but including dynamical friction.}
\label{fig_petsch_1}
\end{figure}

\section{Method}
Finding interaction parameters means solving a minimization problem in an high dimensional parameter space. 
We use a GA based on PIKAIA \citep{charbonneau95} and an improved restricted N-body code based on Toomre \& Toomre \cite{toomre72}.
One important step of improvement was the implementation of static (DM-)halos (e.g.\,isothermal, NFW or Burkert profiles) instead of point masses for galaxies. 
The evolution of our galaxies in time was changed from solving the Keplerian orbit to a time dependent evolution of (overlapping) potentials.
Test particles mimicking the disk are introduced and serve as imprints for the tidal features compared with observed galaxy's HI features.
Recently, we have implemented dynamical friction to the code to allow for later stages of interaction and merging \cite{petsch08}.

\section{Structural parameters of DM haloes}
Our code was already successfully applied to derive tidal models for NGC\,4449 \cite{theis01}, M51 \cite{theis03} an the Magellanic Clouds \cite{Ruzicka09}.
Here we want to focus on the latter and also on the effects of the DM halo and dynamical friction.

The size (and shape) of the DM halo has an important influence to the deflection from a Keplerian orbit and tidal features of interacting disk galaxies.
By comparing observed features with our models, we can derive these halo parameters. 
Fig.\,\ref{fig_petsch_1} (left) shows the large change in deflection angle and tidal features depending on the halo extension, where the final distance keeps almost unaffected.
The latter changes when dynamical friction (accounting for density gradients, mass and distance) is switched on (Fig.\,\ref{fig_petsch_1}, right).

\subsection{The Magellanic Stream and the DM halo of the Milky Way}
\begin{figure}[h]
\centering
\includegraphics[angle=90, width = 8.0cm, clip]{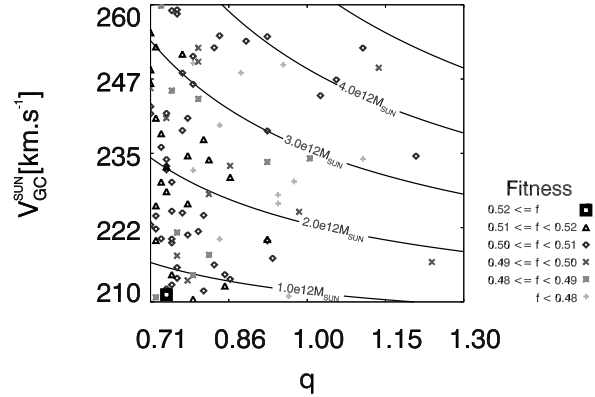}
\caption{Distribution of 100 GA fits of the Magellanic System over the analyzed ranges of the flattening $q$ of the logarithmic dark matter halo of the Galaxy,
and of the solar galactocentric circular velocity $V^\mathrm{sun}_\mathrm{GC}$. The plot also shows the isolines of the total mass of the Milky Way dark matter halo enclosed within
the radius of 250\,kpc. The mass of the logarithmic dark matter halo is a function of 3 independent parameters out of which the flattening $q$ and the solar galactocentric
circular velocity $V^\mathrm{sun}_\mathrm{GC}$ were considered free parameters.
}
\label{fig_ruzicka_02}
\end{figure}

The most recent proper motion measurements for the Large (LMC) and the Small (SMC) Magellanic
Clouds \citep{KalliLMC, KalliSMC} put them on significantly larger galactocentric velocities.
We confirmed the failure of the tidal stripping models of the Milky Way (MW)--LMC--SMC
interaction in such a case \cite{Ruzicka09}. That happened due to the interaction timescale reduced as the Clouds' orbital periods
become comparable to the Hubble time. Shattow et~al. \cite{Shattow09} have resolved the orbital issue by increasing the MW circular velocity.
It affected both the LMC/SMC spatial motion and the MW total mass, leading to multiple perigalactic passages for the Clouds.
We have analyzed the parameter space of the Magellanic interaction
by a robust search algorithm (GA) combined with a fast 3\,D N--body model of
the tidal interaction involving a flattened dark matter halo of the Galaxy. The LSR circular velocity was
varied between 210$\,\mathrm{km\,s^{-1}}$ and 260$\,\mathrm{km\,s^{-1}}$ and linked with the LSR galactocentric distance by the
LSR angular rotation rate of 29.45$\,\mathrm{km\,s^{-1}}$ \citep{Reid04}. Under such assumptions the Clouds were able to stay confined
in the MW halo virial radius for well over 4\,Gyr, and the neutral hydrogen (HI) Magellanic Stream~\citep{bruens05}
was reproduced quite well over the entire range of the LSR circular velocities.
\begin{figure}[h]
\centering
\includegraphics[bb=5 30 585 715, angle=90, scale=0.3, clip]{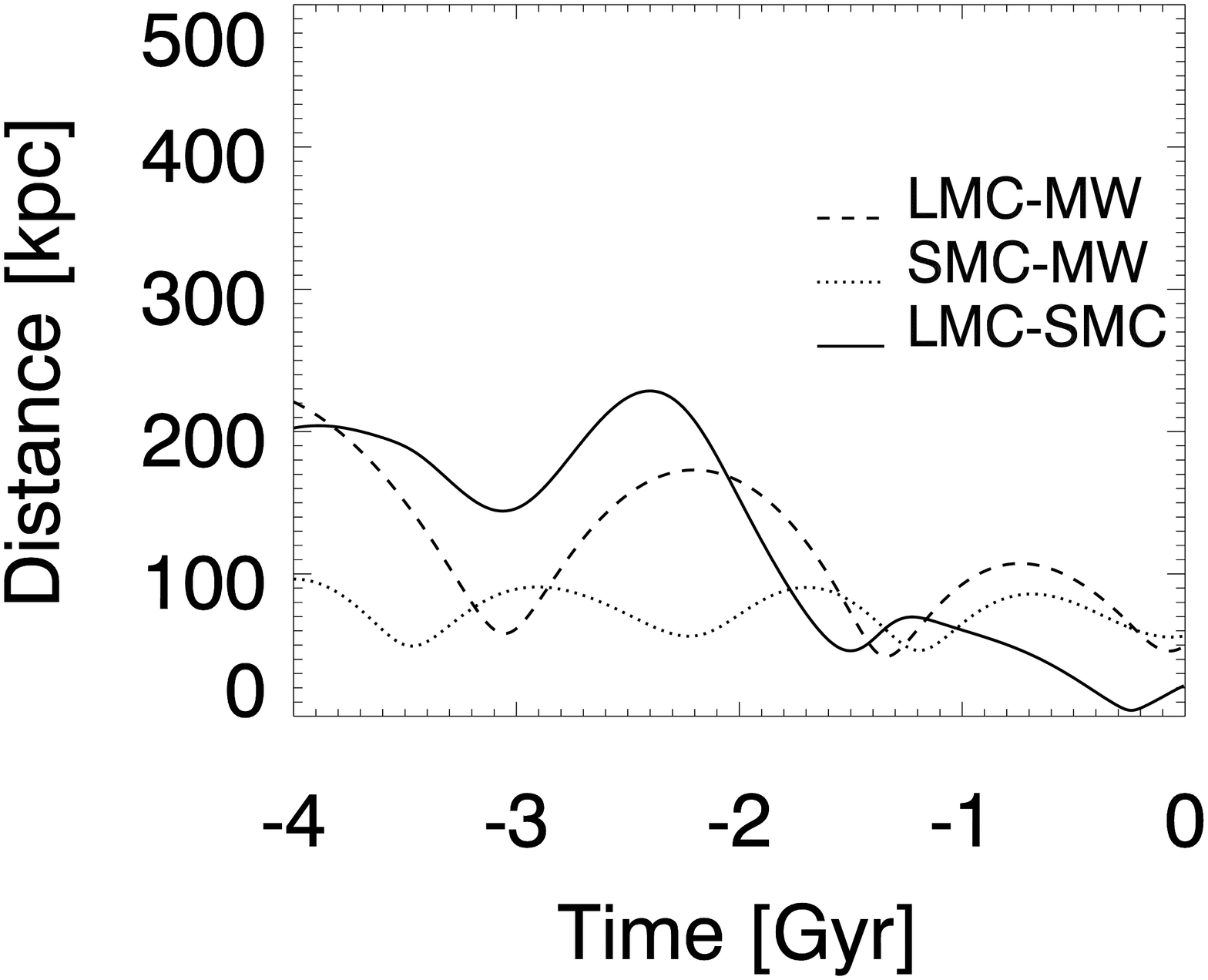}
\includegraphics[bb=10 20 585 755, angle=90, scale=0.3, clip]{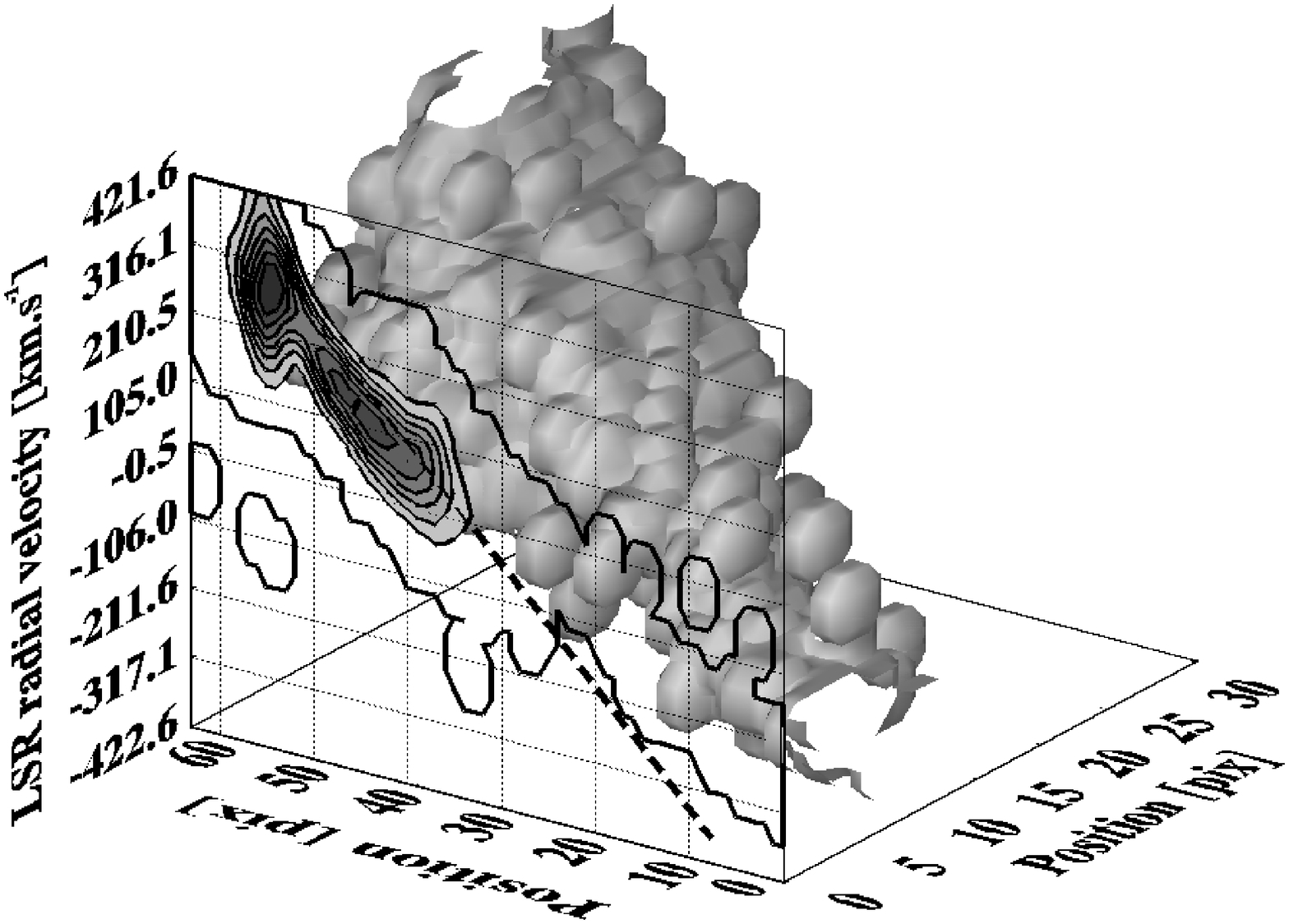}
\caption{Left plot: Orbital evolution of the Clouds over the last 4\,Gyr for a high--quality model with
$V_\mathrm{Sun}=211.2\,\mathrm{km\,s^{-1}}$, $q=0.74$, $M_\mathrm{MW}=0.7\cdot10^{12}\,M_{\odot}$.
Right plot: Visualization of the entire simulated three--dimensional H\,I data cube of the Magellanic System.
The column density isosurface \(\Sigma = 5\cdot10^{-5}\,\Sigma_\mathrm{max}\) is shown,
together with the data cube projected to the 2\,D map of the integrated column density
in the position--radial velocity space.
The dashed line in the position-velocity projections depicts the LSR radial velocity gradient
along the observed Magellanic Stream.}
\label{fig_ruzicka_01}
\end{figure}
The tidal model worked with the new LMC/SMC proper motions over the MW halo mass range of 0.5--3$\cdot\mathrm{10^{12}\,M_{\odot}}$.
In agreement with Ruzicka et~al. \cite{Ruzicka07}, an oblate (flattening$\mathbf{<}$1.0) logarithmic halo was preferred (see Fig.~\ref{fig_ruzicka_02}). However, the modeled
Magellanic Stream was constantly displaced in its projected position compared to the observations.
Introducing the LSR angular rotation rate according to \cite{Reid04} increases the Galaxy mass and reduces the LMC/SMC galactocentric
velocities in their magnitude, and so the tidal model comes back to play.
But it also alters the LMC/SMC velocity vectors in their direction causing an undesired shift
of the Stream in the position--LSR radial velocity space.

The studied parameter space of the Magellanic interaction involves the velocity of the Clouds ($\mu_\mathrm{N}$, $\mu_\mathrm{W}$, $V_\mathrm{rad}$),
LMC/SMC particle disk radii, galactic circular velocity at the solar position ($V_\mathrm{Sun}$),
and the MW halo flattening ($q$). With the use of GA, $\approx 10^6$ parameter combinations were tested in total,
and 100 sets providing satisfactory reproduction of the H\,I Magellanic large--scale structures were collected.





\bibliographystyle{aipproc}   
\bibliography{001ctheis}

\hyphenation{Post-Script Sprin-ger}
\begin{thebibliography}{16}
\expandafter\ifx\csname natexlab\endcsname\relax\def\natexlab#1{#1}\fi
\providecommand{\enquote}[1]{``#1''}
\expandafter\ifx\csname url\endcsname\relax
  \def\url#1{\texttt{#1}}\fi
\expandafter\ifx\csname urlprefix\endcsname\relax\def\urlprefix{URL }\fi
\providecommand{\eprint}[2][]{\url{#2}}

\bibitem[{Holmberg}(1941)]{holmberg41}
E.~{Holmberg}, \emph{ApJ} \textbf{94}, 385 (1941).

\bibitem[{Pfleiderer} and {Siedentopf}(1961)]{pfleiderer61}
J.~{Pfleiderer}, and H.~{Siedentopf}, \emph{Zeitschrift fur Astrophysik}
  \textbf{51}, 201 (1961).

\bibitem[{Toomre} and {To\-omre}(1972)]{toomre72}
A.~{Toomre}, and J.~{To\-omre}, \emph{ApJ} \textbf{178}, 623--666 (1972).

\bibitem[{Zwicky}(1937)]{zwicky37}
F.~{Zwicky}, \emph{ApJ} \textbf{86}, 217 (1937).

\bibitem[{Barnes} and {Hernquist}(1992)]{barnes92}
J.~E. {Barnes}, and L.~{Hernquist}, \emph{ARA\&A} \textbf{30}, 705--742 (1992).

\bibitem[{Charbonneau}(1995)]{charbonneau95}
P.~{Charbonneau}, \emph{ApJS} \textbf{101}, 309 (1995).

\bibitem[{Petsch} and {Theis}(2008)]{petsch08}
H.~P. {Petsch}, and C.~{Theis}, \emph{Astronomische Nachrichten} \textbf{329},
  1046--1049 (2008), \eprint{0810.0625}.

\bibitem[{Theis} and {Kohle}(2001)]{theis01}
C.~{Theis}, and S.~{Kohle}, \emph{A\&A} \textbf{370}, 365--383 (2001),
  \eprint{arXiv:astro-ph/0104304}.

\bibitem[{Theis} and {Spinneker}(2003)]{theis03}
C.~{Theis}, and C.~{Spinneker}, \emph{Ap\&SS} \textbf{284}, 495--498 (2003),
  \eprint{arXiv:astro-ph/0306502}.

\bibitem[{R\accent23u\v{z}i\v{c}ka} et~al.(2009)]{Ruzicka09}
A.~{R\accent23u\v{z}i\v{c}ka}, C.~{Theis}, and J.~{Palou\v{s}}, \emph{ApJ}
  \textbf{691}, 1807--1815 (2009), \eprint{0810.0968}.

\bibitem[{Kallivayalil} et~al.(2006{\natexlab{a}})]{KalliLMC}
N.~{Kallivayalil}, R.~P. {van der Marel}, C.~{Alcock}, T.~{Axelrod}, K.~H.
  {Cook}, A.~J. {Drake}, and M.~{Geha}, \emph{ApJ} \textbf{638}, 772--785
  (2006{\natexlab{a}}), \eprint{arXiv:astro-ph/0508457}.

\bibitem[{Kallivayalil} et~al.(2006{\natexlab{b}})]{KalliSMC}
N.~{Kallivayalil}, R.~P. {van der Marel}, and C.~{Alcock}, \emph{ApJ}
  \textbf{652}, 1213--1229 (2006{\natexlab{b}}),
  \eprint{arXiv:astro-ph/0606240}.

\bibitem[{Shattow} and {Loeb}(2009)]{Shattow09}
G.~{Shattow}, and A.~{Loeb}, \emph{MNRAS} \textbf{392}, L21--L25 (2009),
  \eprint{0808.0104}.

\bibitem[{Reid} and {Brunthaler}(2004)]{Reid04}
M.~J. {Reid}, and A.~{Brunthaler}, \emph{ApJ} \textbf{616}, 872--884 (2004),
  \eprint{arXiv:astro-ph/0408107}.

\bibitem[{Br{\"u}ns} et~al.(2005)]{bruens05}
C.~{Br{\"u}ns}, J.~{Kerp}, L.~{Staveley-Smith}, U.~{Mebold}, M.~E. {Putman},
  R.~F. {Haynes}, P.~M.~W. {Kalberla}, E.~{Muller}, and M.~D. {Filipovic},
  \emph{A\&A} \textbf{432}, 45--67 (2005), \eprint{arXiv:astro-ph/0411453}.

\bibitem[{R\accent23u\v{z}i\v{c}ka} et~al.(2007)]{Ruzicka07}
A.~{R\accent23u\v{z}i\v{c}ka}, J.~{Palou\v{s}}, and C.~{Theis}, \emph{A\&A}
  \textbf{461}, 155--169 (2007), \eprint{arXiv:astro-ph/0608175}.

\end{thebibliography}

\end{document}